\documentclass[aps,prl,showpacs,twocolumn,superscriptaddress]{revtex4-1}

\usepackage{amsmath}
\usepackage{color}

\newcommand{\ket}[1]{|{#1}\rangle}
\newcommand{\bra}[1]{\langle{#1}|}
\newcommand{\T}{\mathop{\text{T}}\nolimits}

\definecolor{dgreen}{rgb}{0,0.5,0}
\definecolor{delete}{cmyk}{0.5,0,0,0}

\begin{document}
\title{Non-Abelian Phases from a Quantum Zeno Dynamics}

\author{Daniel Burgarth}
\affiliation{Institute of Mathematics and Physics, Aberystwyth University, SY23 3BZ Aberystwyth, United Kingdom}

\author{Paolo Facchi}
\affiliation{Dipartimento di Fisica and MECENAS, Universit\`a di Bari, I-70126  Bari, Italy}
\affiliation{INFN, Sezione di Bari, I-70126 Bari, Italy}

\author{Vittorio Giovannetti}
\affiliation{NEST, Scuola Normale Superiore and Istituto Nanoscienze-CNR,  
I-56126 Pisa, Italy}

\author{Hiromichi Nakazato}
\affiliation{Department of Physics, Waseda University, Tokyo 169-8555, Japan}

\author{Saverio Pascazio} 
\affiliation{Dipartimento di Fisica and MECENAS, Universit\`a di Bari, I-70126  Bari, Italy}
\affiliation{INFN, Sezione di Bari, I-70126 Bari, Italy}

\author{Kazuya Yuasa}
\affiliation{Department of Physics, Waseda University, Tokyo 169-8555, Japan}
\date[]{September 18, 2013}

\begin{abstract}
A connection is established between the non-Abelian phases obtained via adiabatic driving and that acquired via a quantum Zeno dynamics induced by repeated projective measurements. In comparison to the adiabatic case, the Zeno dynamics is shown to be more flexible in tuning the system evolution, which paves the way to the implementation of unitary quantum gates and applications in quantum control.  
\end{abstract}


\maketitle

\textit{Introduction.---}
The possibility of engineering arbitrary dynamics by modifying the system Hamiltonian in real time is the main paradigm of control theory. 
For a quantum system it is known that highly nontrivial evolution  might be induced by exploiting the interference effects arising when the system is driven along special  loops in the space of the control parameters. 
These techniques are at  the core of geometric~\cite{EKERT} and topological quantum computation~\cite{PACHOS}. 
The main drawback of the adiabatic  approach is associated with the fact that 
 the realization of gates requires in general a long running time, 
 in order to minimize nonadiabatic transitions.   
 Even though in some cases this limitation can be overcome through a clever design of the control pulses (e.g., see Refs.\ \cite{CONTROL1,CONTROL2}),  this typically entails extra costs which might conflict with the resources at our disposal.

A completely different approach to control is the possibility of exploiting the quantum Zeno effect and dynamics~\cite{ref:ZenoSubspaces}. 
The key idea is to engineer a given dynamical evolution by a rapid sequence of projections \cite{vN,AA}. 
This yields in general a Berry phase \cite{ref:FKPS}. This evolution has been studied in the literature and is sometimes referred to as anti-Zeno effect
\cite{ref:antiZeno,ref:Diosi}. We intend to utilize efficiently these ideas in order to engineer unitary holonomic gates through loops in parameter space  \cite{ZANNA3,ZANNA4,ref:DuanCiracZoller,FFFGP}. We notice that a gate hinging upon such a Zeno mechanism can be very fast \cite{ref:FKPS}, its speed depending essentially on how fast the projections can be performed in practice. 

Starting from the seminal work by Franson \textit{et~al.}\ \cite{ZENOGATE1} several  attempts to exploit these ideas have been discussed in  quantum optics \cite{ZENOGATE2,ZENOGATE3,ZENOGATE4,ZENOGATE5} with the aim of inhibiting the failure events that would  otherwise occur in a linear optics approach to quantum computing. More recently, control induced via Zeno dynamics has been also analyzed in the broader context of quantum error correction \cite{LIDAR1,LIDAR2} and quantum computation \cite{LANDAHL,DORIT}.

In the present work we unearth a connection between adiabatic driving and quantum Zeno effect. We show that the implementation of 
evolution through measurements provides additional freedom in tuning the system dynamics which paves the way to new control applications. Finally, we demonstrate how a holonomic evolution can be achieved entirely through dissipation engineering.

\textit{Adiabatic dynamics.---}%
As a starting point of our discussion it is convenient to briefly recall basic facts about adiabatic evolutions \cite{MESSIAH,REV1,REV2,REV3} and their application
to quantum control and quantum computation \cite{EKERT,ZANNA3,ZANNA1,ZANNA4}.
Consider a time-dependent Hamiltonian $H(t)$, which is diagonalizable at any instant $t$ as
\begin{equation}
H(t)= \sum_n E_n(t) P_n(t),\ \ 
E_n(t)\neq E_{n'}(t)\ \ (n\neq n'),
\label{eqn:InstE}
\end{equation}
where $P_n(t)$ are orthogonal projections onto the instantaneous eigenspaces $\mathcal{S}_n(t)$ of $H(t)$, that are in general multi-dimensional, and $E_n(t)$ are the associated energies.
The parametric temporal dependence of $P_n(t)$ is determined  by a continuous  family of unitary operators $W(t)$ which define the mapping
\begin{equation}
P_n(t)=W(t)P_n(0) W^\dag(t),\ \ \forall n,\quad W(0)=\openone,
\label{eqn:W}
\end{equation}
where $W(t)$ need not form a one-parameter group.
Factoring out $W(t)$ and the dynamical phases from the evolution of the system $U(t)$, generated by $H(t)$, 
as ($\hbar=1$) 
\begin{equation}
U(t) =W(t)\sum_{n}e^{-i\int_0^tds\,E_n(s)}P_n(0) {U}_G(t),
\end{equation}
the unitary operator 
${U}_G(t)$, with Hamiltonian $H_G(t)$, is responsible for 
the inter- and intra-couplings  between the energy eigenspaces $\mathcal{S}_n(t)$.
Under the adiabatic approximation [see (\ref{ADCOND})  below]
the transitions across the energy eigenspaces $\mathcal{S}_n(t)$ are dynamically suppressed, and one ends up with $U_G(t)\simeq U_G^\text{(diag)}(t):=\prod_nU_G^{(n)}(t)$, where 
$U_G^{(n)}(t)$ is generated by
\begin{equation}
{H}^{(n)}_G(t) := i   P_n(0) \dot{W}^\dag(t)W(t)P_n(0).
\label{eqn:HCint}
\end{equation}
The dynamics is still capable of inducing nontrivial rotations
  within individual subspaces $\mathcal{S}_n(t)$,
   in the form 
of non-Abelian Berry phases (holonomies)  that possess a distinctive  geometrical~\cite{NOTA1} or, in some special cases, topological  character (see below for an explicit example). 
Specifically, assuming that at time $t=0$ the system is initialized in a state $|\psi(0)\rangle$ belonging to the $n$th subspace $\mathcal{S}_{n}(0)$, 
the state of the system at time $t\ge0$ is described by the vector
\begin{equation}
|\psi(t)\rangle \simeq  e^{-i\int_0^tds\,E_n(s)}W(t) {U}^{(n)}_G(t)|\psi(0)\rangle,\label{ex1}
\end{equation} 
which remains in the $n$th subspace $\mathcal{S}_n(t)$ 
[notice that the projection $P_n(0)$ has
been absorbed in $|\psi(0)\rangle$ exploiting the fact that it commutes with ${U}^{(n)}_G(t)$].
Even if the Hamiltonian $H(t)$ is driven back to the original one $H(0)$  after the slow driving and the system comes back to the original subspace $\mathcal{S}_n(0)$, the state of the system $|\psi(t)\rangle$ can be different from the initial one $|\psi(0)\rangle$, rotated according to (\ref{ex1}).
This rotation can be used to realize gate operations for quantum computation, which can have high stability due to their geometric/topological nature.

Note that the adiabatic evolution (\ref{ex1}) is an approximation, ensured as long as 
 the probability $q_n(t)= \bra{\psi(t)}(1-P_n(t))\ket{\psi(t)}$ of the system escaping from the $n$th subspace $\mathcal{S}_n(t)$ is negligible, i.e.,
\begin{eqnarray}
  q_n(t) &=& \bra{\psi(0)} \tilde{U}_G^\text{(off)\dag}(t) (1-P_n(0)) \tilde{U}_G^\text{(off)}(t) \ket{\psi(0)}
  \nonumber\\
&\simeq& \bra{\psi(0)} \left( \int_0^t d s \,\tilde{H}^\text{{(off)}}_G(s)\right)^2\ket{\psi(0)} \ll  1,  \label{ADCOND}
\end{eqnarray}
where 
$\tilde{H}_G^\text{(off)}:=U_G^\text{(diag)\dag}[H_G-\sum_nH_G^{(n)}]U_G^\text{(diag)}$ is  the generator of $\tilde{U}_G^\text{(off)}$, responsible for the escape.

\textit{Zeno dynamics.---}%
Let us next analyze the dynamics of a quantum system with a Hamiltonian $H_0(t)$ [not necessarily the same as $H(t)$ in (\ref{eqn:InstE})], under a sequence of projective measurements described by a set of time-dependent projections $P_n(t_k)$ performed at $t_k=kt/N$ ($k=0,\ldots,N$), where the measurement is \emph{changed} step by step as in (\ref{eqn:W}), unlike in the ordinary Zeno effect \cite{ref:ZenoSubspaces,
ref:ZenoBB,ref:DecoCont}.
Suppose then  that, starting from an initial state $|\psi(0)\rangle$ in the $n$th subspace  $\mathcal{S}_n(0)$, as in the case of (\ref{ex1}), the system is found to remain in its rotated counterparts $\mathcal{S}_n(t_1),\mathcal{S}_n(t_2),\ldots$ by the $N$ successive measurements.
Under this condition, the evolution of the system is  described  by the operator
\begin{align} 
V^{(n)}_N(t) 
&:=\prod_{k=0}^{N-1} P_n(t_{k+1})U_0(t_{k+1},t_k)P_n(t_k)
\nonumber\\
&= W(t) \prod_{k=0}^{N-1}P_n(0) \tilde{U}_0(t_{k+1},t_k)P_n(0),
\label{eqn:ZenoEvo}
\end{align}
where $U_0(t,t'):=\T e^{-i\int_{t'}^tds\,H_0(s)}=W(t)\tilde{U}_0(t,t')W^\dag(t')$, and the products are understood to be time-ordered  [with
later times (larger $k$) to the left] hence forth.
Seen from the frame rotating according to $W(t)$, the system appears to evolve with $\tilde{U}_0(t,t')$, during which it is repeatedly measured with fixed projections $P_n(0)$.
With these definitions the state of the system at time $t$ is given by 
$|\psi(t)\rangle = V^{(n)}_N(t) |\psi(0)\rangle/\sqrt{p_N^{(n)}(t)}$, where 
\begin{equation}
p_N^{(n)}(t) : =\|  V^{(n)}_N(t) |\psi(0)\rangle\|^2
\end{equation} 
is the probability of the realization of the conditional evolution.
For large $N$, each factor in the projected evolution operator (\ref{eqn:ZenoEvo}) reduces to
\begin{equation}
P_n(0) \tilde{U}_0(t_{k+1},t_k)P_n(0) 
= e^{-i {H}^{(n)}_Z(t_k)\frac{t}{N}}P_n(0) + O(t^2/N^2),
\end{equation}
where  ${H}^{(n)}_Z(t)$ is an emergent Hamiltonian given by
\begin{equation}
{H}^{(n)}_Z(t)
= P_n(0) W^\dag(t)H_0(t)W(t) P_n(0) +  H^{(n)}_G(t),
\label{eqn:HZn}
\end{equation}
with $ H^{(n)}_G(t)$  being the adiabatic Hamiltonian (\ref{eqn:HCint}). 
From the point of view of differential geometry, if one neglects the Zeno-projected part of the Hamiltonian, this yields nothing but the connection (vector potential) of the curvature (Yang-Mills field). A similar expression was derived by Anandan and Aharonov for the Abelian case \cite{AA}, while non-Abelian holonomies obtained through sequences of projective measurements were discussed by Anandan and Pines \cite{AP}, and later in Refs.\ \cite{CSV,OC} in the context of open system dynamics.

We are mostly interested in the interplay between the two components of the Hamiltonian (\ref{eqn:HZn}).
Such interplay is typical of the Zeno mechanism (where it is called Zeno dynamics \cite{ref:ZenoSubspaces}) and, as we shall see, makes it possible to speed up physical operations. 

In the Zeno limit $N\to\infty$ keeping $t$ finite,  one gets
\begin{align}
V_N^{(n)}(t)
&\sim W(t)
\prod_{k=0}^{N-1}
e^{-i {H}^{(n)}_Z(t_k)\frac{t}{N}}P_n(0)
\nonumber\\
&\to W(t){U}^{(n)}_Z(t)P_n(0)  ,
\label{zeno}
\end{align}
with 
\begin{equation}
{U}^{(n)}_Z(t) := \T e^{-i\int_0^t ds\,{H}_Z^{(n)}(s)}
\end{equation} 
being the Zeno unitary transformation, which maps $\mathcal{S}_n(0)$ into itself. 
Accordingly $p_N^{(n)}(t)  \rightarrow 1$, i.e., the system remains in $\mathcal{S}_n(t)$ at all $t$ with certainty, while it is rotated there as 
\begin{equation}
|\psi(t) \rangle= W(t) {U}^{(n)}_Z(t)|\psi(0)\rangle.\label{ex2}
\end{equation} 
This result bears a striking resemblance with the adiabatic evolution in (\ref{ex1}) with (\ref{eqn:HCint}). 
In particular, by choosing $H_0(t)= H(t)$ and performing the measurements such that the associated projections $P_n(t)$ coincide with the instantaneous eigenprojections of $H(t)$ in (\ref{eqn:InstE}), one gets 
${H}^{(n)}_Z(t)= E_n(t)P_n(0) +{H}^{(n)}_G(t)$, which exactly reproduces (\ref{ex1}). 

It should be noticed however that the  correspondence  between the adiabatic and Zeno scenarios holds since we considered as 
the initial state of the system  a vector $|\psi(0)\rangle$  contained in a \textit{single}   subspace $\mathcal{S}_n(0)$. 
 This is a crucial assumption since
 due to the projective measurements the Zeno procedure (\ref{eqn:ZenoEvo}) naturally  leads to the decoherence of any superposition 
 present initially across subspaces, while the adiabatic evolution preserves it. 
Specifically for a generic initial state $\rho(0)$ inhabiting across different subspaces $\mathcal{S}_n(0)$,
 the Zeno dynamics under a sequence of frequent \textit{nonselective} measurements 
 \begin{equation}
\label{eq:dephasing}
{\cal P}(t)\rho := \sum_n P_n(t) \rho P_n(t),
\end{equation} 
which provides a dephasing channel removing all the correlations among the subspaces $\mathcal{S}_n(0)$, yields
\begin{equation}
\rho(t)=\mathcal{W}(t)\mathcal{U}_Z(t)\mathcal{P}(0)\rho(0), \label{eqgen}
\end{equation} 
instead of (\ref{ex2}), where  $\mathcal{U}_Z(t)\rho=U_Z(t)\rho U_Z^\dag(t)$ with $U_Z(t) = \prod_n U_Z^{(n)}(t)$, and $\mathcal{W}(t)\rho=W(t)\rho W^\dag(t)$.
Nonetheless, as long as we are interested in the dynamical processes
 taking place \textit{inside} a given subspace $\mathcal{S}_n(t)$ the equivalence between the two descriptions  is guaranteed.

\textit{Controls and constraints.---}%
Equation (\ref{zeno}) with (\ref{eqn:HZn}) shows that sequences of time-dependent projective measurements aimed at checking whether or not a system  initially prepared in $\mathcal{S}_n(0)$ remains in the $n$th subspace $\mathcal{S}_n(t)$ enable one to reproduce the evolution attainable by enforcing the adiabatic dynamics (\ref{ex1}). 
Interestingly enough, however, while the latter requires time scales  which determine the variations of the driving $W(t)$ to be long with respect to the inverse of the minimum gap of the energy spectrum of $H(t)$ along the trajectory, the mapping (\ref{ex2}) is free from this constraint: the adiabatic evolution can in principle be realized with no speed limitation.
In the Zeno scenario, on the other hand, condition (\ref{ADCOND}) is replaced by the Zeno limit (\ref{zeno}), which requires  the projective  measurements varied continuously in time to be performed frequently enough at a sufficiently rapid pace. In other words, moving from (\ref{ex1}) to (\ref{ex2}) we trade the \textit{slow-driving} requirement \cite{LANDAHL,DORIT} for the adiabatic evolution with the \textit{fast-monitoring}
requirement for the Zeno paradigm.

Another remarkable difference between the two procedures is that
in the Zeno case, once fixed the parametric dependence of $P_n(t)$, we still have the freedom to choose $H_0(t)$ in order to design ${H}^{(n)}_Z(t)$
[this  freedom being absent in the adiabatic scenario, where the dynamical part $P_n(0) W^\dag(t)H(t)W(t) P_n(0) = E_n(t) P_n(0)$ is diagonal and automatically determined by 
the given Hamiltonian $H(t)$]. As a matter of fact Eq.\ (\ref{eqn:HZn}) can be expressed as 
 \begin{equation}
{H}^{(n)}_Z(t) \label{new1}
=  i   P_n(0) \dot{\tilde{W}}^\dag(t)\tilde{W}(t)P_n(0)
\end{equation}
by introducing the unitary  operator 
\begin{equation}
\tilde{W}(t) =U_0^\dag(t,0)W(t). 
\end{equation}
This expression reveals that $H_0(t)$ plays the role of effectively modifying the way of rotating the measurement basis (namely the subspaces followed by the system during the Zeno procedure) from $W(t)$ to $\tilde{W}(t)$.
In view of these considerations we identify  two configurations that deserve special attention:

(i) No Hamiltonian $H_0(t) =0$. In this case the system does not possess any intrinsic dynamics. Still, a nontrivial unitary evolution (\ref{ex2}) is induced via the Zeno procedure, which simulates the adiabatic evolution with a Hamiltonian $H(t)$, apart from the phase in each subspace. Conceptually this is reminiscent of what happens in one-way quantum computation \cite{one-way}, where an effective temporal evolution  is introduced via sequences of properly selected measurements performed on an otherwise static quantum register.  Notably, the Zeno limit (\ref{zeno}) does not pose
  constraints on the speed at which the instantaneous  measurements have  to be varied in time (no other time scale being present in the system): it only requires  a continuity condition [i.e., the  projections $P_n(t_k)$ performed at the $k$th step must be close to those $P_n(t_{k+1})$ performed at the $(k+1)$th step].

(ii) A constant Hamiltonian $H_0(t)=H_0$.  
In this  case the system is characterized by a proper intrinsic (uncontrolled) Hamiltonian,  
which sets the pace at which the measurements have to be performed in order to ensure the Zeno limit (\ref{zeno}). Once this limit is reached
the resulting evolution of the system can be forged along trajectories which could not be realized by $H_0$. 
For instance, a real life implementation of the {\it  wagon-wheel optical illusion  effect} can be induced on the system by taking 
$W(t) = e^{- 2 i H_0t}$. Under this condition Eq.\ (\ref{eqn:HZn}) yields $H_Z^{(n)}=- P_n(0)  H_0 P_n(0)$
resulting in an effective time reversal of the dynamics induced by $H_0$. 

\textit{Example.---}%
The possibility of exploiting  the correspondence between (\ref{ex1}) and (\ref{ex2})  paves the way to a new form of quantum control, where frequent measurements
are introduced to replace the adiabatic driving or, even better, 
to compensate possible departures from the adiabatic regime, leading hence to a speed-up of the resulting  transformation. 
To clarify this point we find it useful to consider the following simple model, where
 a three-level system $\{\ket{1},\ket{2},\ket{3}\}$ is driven with a Hamiltonian $H(t)$ depending on two time-dependent real parameters $a(t)$ and $b(t)$,\begin{equation}
H=\begin{pmatrix}
\sqrt{a^2+b^2}&a&b\\
a&\frac{a^2}{\sqrt{a^2+b^2}}&\frac{ab}{\sqrt{a^2+b^2}}\\
b&\frac{ab}{\sqrt{a^2+b^2}}&\frac{b^2}{\sqrt{a^2+b^2}}
\end{pmatrix}.
\end{equation}
This admits two instantaneous eigenvalues, which in terms of the polar coordinates $a=r\cos\theta$ and $b= r \sin\theta$ 
are given by $E_0=0$ and $E_1=2r$, 
with the former being twofold degenerated and the gap between the two eigenvalues closing at the critical point $(a,b)=(0,0)$.  The corresponding instantaneous eigenspaces $\mathcal{S}_0(
\theta)$ and 
$\mathcal{S}_1(\theta)$ are specified by the projections
$P_0(\theta)=\ket{E_0(\theta)}\bra{E_0(\theta)}+\ket{E_-(\theta)}\bra{E_-(\theta)}$ and $P_1(\theta)=\ket{E_+(\theta)}\bra{E_+(\theta)}$
with 
\begin{equation}
\begin{cases}
\ket{E_\pm(\theta)}
=\frac{
\ket{1}\pm(\cos\theta\,\ket{2}+\sin\theta\,\ket{3})
}{\sqrt{2}}
=W(\theta)\ket{E_\pm(\theta_0)},\\
\ket{E_0(\theta)}
=-\sin\theta\,\ket{2}+\cos\theta\,\ket{3}
=W(\theta)\ket{E_0(\theta_0)},
\end{cases}
\!\!\!\!\!\!
\end{equation}
where $W(\theta)=e^{-iG(\theta-\theta_0)}$ with $G=-i\ket{2}\bra{3}+i\ket{3}\bra{2}$ is the unitary transformation inducing the parametric rotations of the eigenspaces from $\theta_0:=\theta(0)$.
For this model the adiabatic regime (\ref{ADCOND}) is guaranteed when \begin{equation}
|\dot{\theta}(t)| \ll r(t)
\end{equation}
for all $t$ along the trajectory in the $(a,b)$ plane [this of course excludes the possibility that the trajectory passes through the critical point $(a,b)=(0,0)$, where $r$, and hence the instantaneous gap, vanishes].
In the subspace $\mathcal{S}_0(\theta)$,  the  generator of the adiabatic evolution (\ref{eqn:HCint}) 
is given by
${H}_G^{(0)}(t)
=-G_0\dot{\theta}(t)/\sqrt{2}$
with 
\begin{equation}
G_0
=-i  \ket{E_0(\theta_0)} \bra{E_-(\theta_0)} +i  \ket{E_-(\theta_0)} \bra{E_0(\theta_0)},
\end{equation}
yielding the adiabatic unitary gate
${U}^{(0)}_G(t) =e^{\frac{i}{\sqrt{2}}G_0 [ \theta(t) - \theta_0]}.$
This transformation, besides possessing a geometrical character,  has also a  topological character. Indeed for any closed loop in the parameter space 
 one gets  
 \begin{equation}
{U}_G^{(0)} = e^{i \Delta m \sqrt{2}\,\pi G_0}
\end{equation} 
with $\Delta m$ being the difference between the number of anti-clockwise and clockwise  windings around the critical point $(a,b)=(0,0)$ (if the loop does not encircle the origin then $\Delta m=0$). 
Following our analysis the same evolution  can be induced by the Zeno procedure (\ref{ex2}), operating on a 3-level system with no intrinsic dynamics (i.e., $H_0=0$) and performing sequences of instantaneous measurements which check whether the system belongs to the 
 subspace $\mathcal{S}_0(\theta_0),\mathcal{S}_0(\theta_1),\mathcal{S}_0(\theta_2),\ldots$\@

Notice that, along the lines discussed in point (ii) of the previous section, the above rotation angle $\Delta m \sqrt{2}\,\pi$ after $\Delta m$ windings of  Zeno driving can be modified by applying a Hamiltonian 
\begin{equation}
H_0(t)=\alpha G\dot{\theta}(t)
\end{equation} 
during the operation.
In this case  Eq.~(\ref{eqn:HZn}) yields $H_Z^{(0)}(t)=-(1-\alpha)G_0\dot{\theta}(t)/\sqrt{2}$, and one gets 
\begin{equation}
{U}_Z^{(0)} = e^{i(1-\alpha)\Delta m\sqrt{2}\,\pi G_0}.
\end{equation}

\emph{Zeno gate by dissipation.---}%
There are different implementations of the Zeno dynamics \cite{ref:ZenoSubspaces,
ref:ZenoBB,ref:DecoCont}: via projective measurements studied above, via unitary kicks (including the bang-bang control), and via a strong coupling to an external agent.
All these strategies are applicable for inducing geometric phases and for realizing Zeno gates by the procedure detailed above.
More interestingly, dephasing induced by an external environment can be utilized to implement the Zeno gates. This possibility is allowed by the peculiar fact that,
when monitoring the system on a rapid pace, we are not required to read out the results of the measurements. As a matter of fact, the whole procedure is 
explicitly designed in such a way that  the measurements are expected to give always the same result (i.e., the system is always found inside the $n$th subspace).
Given that, it makes absolutely no difference if we let the environment to perform the projections $P_n(t)$. To see how this works explicitly, suppose that our system
is evolving through the following time-dependent master equation,
\begin{equation}
\dot{\rho}(t)={-i}[H_0(t),\rho(t)]
+\mathcal{L}(t)\rho(t),
\label{eqn:MasterEq}
\end{equation}
where 
\begin{equation}
\mathcal{L}(t)\rho=-\frac{1}{2}\gamma[L^2(t)\rho+\rho L^2(t)-2L(t)\rho L(t)]
\end{equation}
with 
\begin{equation}
L(t)=\sum_n\alpha_nP_n(t), \qquad (\alpha_n\neq\alpha_{n'}\; \text{for}\;  n\neq n').
\end{equation}
Going to the rotating frame defined by $\tilde{\rho}(t)=W^\dag(t)\rho(t)W(t)$, the master equation (\ref{eqn:MasterEq}) is converted into
\begin{equation}
\dot{\tilde{\rho}}(t)
={-i}[\tilde{H}(t),\tilde{\rho}(t)]
+\mathcal{L}\tilde{\rho}(t),
\label{eqn:MasterEqR}
\end{equation}
where $\tilde{H}(t) = W^\dag(t) H_0(t) W(t) + i \dot{W}^\dag(t) W(t)$ and $\mathcal{L}=\mathcal{L}(0)$.
Note also that 
$e^{\mathcal{L} t} \xrightarrow{t\to\infty}\mathcal{P}$,
with $\mathcal{P}=\mathcal{P}(0)$ being the projection~(\ref{eq:dephasing}).
Therefore, taking $\gamma\to\infty$ in the Dyson series of~(\ref{eqn:MasterEqR}), one gets
\begin{align}
\tilde{\rho}(t)
&=e^{\mathcal{L}t}\rho(0)
-i\int_0^tdt'e^{\mathcal{L}(t-t')}
[\tilde{H}(t'),e^{\mathcal{L}t'}\rho(0)]+\cdots
\nonumber\\
&\to
\mathcal{P}\rho(0)
-i\int_0^tdt'\,\mathcal{P}
[\tilde{H}(t'),\mathcal{P}\rho(0)]
+\cdots,
\end{align}
which, exploiting the fact that ${\cal P}\tilde{H}(t)= \sum_n H_Z^{(n)}(t)=H_Z(t)$,
 obeys a von Neumann equation
\begin{equation}
\dot{\tilde{\rho}}(t)
=-i[H_Z(t) ,\mathcal{P}\tilde{\rho}(t)],\quad
\tilde{\rho}(0)=\mathcal{P}\rho(0).
\label{eqn:ZenoMaster}
\end{equation}
Integrating it and  moving back to the canonical reference frame, this yields
  the solution (\ref{eqgen}) as anticipated. 

\textit{Conclusions.---}%
We have provided a formal connection between the adiabatic theorem and Zeno dynamics and shown that the latter provides more flexibility in implementing gates with topological character. While a detailed analysis of the resources involved is beyond the scope of this paper, it is clear that our result opens an avenue for quantum control techniques
based on a continuous  monitoring of the system dynamics. Accordingly a given target  quantum evolution is induced via a sort of stroboscopic approach in which the system of interest is projected  onto subspaces that are externally steered.


\begin{thebibliography}{10}
\bibitem{NOTA1}
For instance assuming a periodic evolution of $W(t)$ the unitary $U_G(t)$ after
  a cycle depends only on the surface encircled by the trajectory on the space
  of the control parameters which govern the modulation.
\end{thebibliography}

\end{document}